# Interplay between magnetism and magnetocaloric effect in NdCuSi

Sachin Gupta[1], K. G. Suresh[1,*], and A. K. Nigam[2]

[1]*Department of Physics, Indian Institute of Technology Bombay, Mumbai-400076, India*

[2]*Tata Institute of Fundamental Research, Homi Bhabha Road, Mumbai-400005, India*

## Abstract

We report the magnetic, thermal, magnetocaloric and transport properties of polycrystalline NdCuSi. The compound shows a weak antiferromagnetic ordering, as confirmed by different measurements. Arrott plots indicate that the magnetic phase transition is found to be second order in nature. The isothermal magnetic entropy change $(\Delta S_M)$ calculated using the magnetization data gives a value of 11.1 J/kg K for a field change of 50 kOe. The $\Delta S_M^{max}$ follows $h^{2/3}$ dependence (h is the reduced field) with a negative intercept, which is predicted by mean field theory for second order phase transitions. Landau theory of phase transition was applied to calculate change in entropy, which was found to be in good agreement with that calculated directly from the magnetization data. Temperature dependence of electrical resistivity shows a pronounced anomaly at the ordering temperature, which is attributed to the spin fluctuations.

## Introduction

Rare earth (R) based intermetallic compounds RTX (T= transition metal and X= p-block element) are found to show a variety of magnetic, electrical and thermal properties, mainly due to the difference in the crystal structure among the members. These compounds have been studied for their fundamental magnetic properties and only very few attempts have been made to estimate their application potential. It is observed that compounds in RCuSi series, belonging to the RTX family, are isostructural and crystallize in the hexagonal crystal structure. R, T, and X atoms in this series occupy 2a, 2c, and 2d sites, respectively. Some authors [1-4] have reported that two types of structures are possible for the RCuSi series; one is ZrBeSi type (SG P $6_3$/mmc) and the other is $AlB_2$ type (SG=P6/mmm). It is also known that most of compounds in this series are antiferromagnetic. CeCuSi is reported to be ferromagnetic below 15.5 K [5]. DyCuSi [6] and HoCuSi [7] are found to exibit giant magnetocaloric effect (MCE) with isothermal magnetic entropy changes (-$\Delta S_M$) of 24 and 33.1 J/kg K, respectively for a field change of 50 kOe. Motivated by the results on these two compounds, we have studied the magnetic and magnetocaloric properties of some members of this series and in this paper, we report the results obtained in NdCuSi. The magnetic, thermal, magnetocaloric and transport properties of this compound are discussed. The compound shows weak antiferromagnetic ordering at low temperatures, which is broken by applied field. Magnetic properties of the compound are correlated with the magnetocaloric properties.

## Experimental Details

NdCuSi compound was prepared by arc melting of the constituent elements with purity better than 99.9 % in a water cooled copper hearth under argon atmosphere using titanium as a oxygen getter. The alloy was melted six times to make the composition homogeneous. It was then sealed in evacuated quartz tube and annealed for 10 days at 850℃. The phase purity of annealed sample was examined by the analysis of x-ray diffraction (XRD) pattern taken from X'Pert Pro diffractometer at room temperature. Magnetization M(T,H), heat capacity C(T) and resistivity (ρ) measurements were carried out in a Quantum Design, Physical Property Measurement System (PPMS-6500). Thermal relaxation method was used to carry out the heat capacity measurement. The resistivity was measured by employing standard four probe technique applying an excitation current of 150 mA.

**Results and Discussion**

The Rietveld refinement of room temperature powder XRD pattern shows no detectable impurity. It is found that the compound crystallizes in the hexagonal crystal structure with the space group P 6$_3$/mmc (SG#194). The lattice parameters obtained from the least square fit are found to be a=b=4.2103(1) A°, c=7.8289(3) A°. These values are very close to the ones reported [4]. The XRD pattern with Rietveld refinement is shown in Fig. 1.

Temperature dependence of dc magnetic susceptibility ($\chi$) and inverse magnetic susceptibility are shown in Fig.2. The magnetic transition at low temperatures is not clear in main plot, but can be seen in the inset of Fig. 2. The cusp-like nature of $\chi$ at low temperatures indicates the antiferromagnetic ordering. The Curie-Weiss law, $\chi^{-1} = (T - \theta_p)/C_m$ ($C_m$ is the molar Curie constant and $\theta_p$ is paramagnetic Curie temperature) yields the effective magnetic moment ($\mu_{eff}$) of 3.87 $\mu_B$/f.u. and the paramagnetic Curie temperature ($\theta_p$) of -11 K. The value of $\mu_{eff}$ is close to that of free $Nd^{3+}$ ion and the negative sign of $\theta_p$ suggests the antiferromagnetic nature of this compound.

Fig. 3(a) displays the field dependence of magnetization in the temperature range of 2-60 K with a temperature interval of 2 K and for fields up to 50 kOe. The field sweep rate in magnetization measurement was set low enough to get the magnetization curves in the isothermal manner. Careful observation of Fig. 3(a) reveals that the compound shows very small magnetic hysteresis at 2 K, which disappears at higher temperatures. The other point to be noted from the magnetization isotherms is that the difference in the magnetization between two temperatures is very large, which reflects a large magnetic entropy change near the ordering temperature. The inset of Fig. 3(a) shows the M vs. H plots at 2, 3 and 4 K with a linear fit to the magnetization at 2 and 4 K. Magnetization data was not fitted for 3 K for the sake of clarity. From this inset one can note two points: (i) the magnetization at 3 K is less than that at 2 K and (ii) the curve is linear only below 1 kOe, after which it starts deviating from linearity. These two points manifest a weak antiferromagnetic interaction in this compound and reveal the AFM-FM ordering above 1 kOe.

In order to find out the order of magnetic phase transition, we evaluated the Arrott Plots ($M^2$ vs. H/M), shown in Fig. 3(b). Banerjee criterion [8] suggests that the positive slope of all the

Arrott plots of a compound gives second order magnetic phase transition while negative slope of some of these plots indicates the first order nature. In the present case, as the plots do not show any negative slope, it is clear that the compound shows second order magnetic phase transition.

To get a better understanding of this compound, the heat capacity measurement was done in the temperature range of 2-100 K and at 0, 10, 20, 50 kOe fields (Fig. 4). The zero field heat capacity data shows a λ-like peak at the onset of magnetic ordering and confirms the second order magnetic transition, as suggested by the Arrott plots. From the lower inset of Fig. 4, one can see that on increasing the field, the peak shifts to higher temperatures and proves that the application of field suppresses the antiferromagnetic interaction. The upper inset shows the fit of Debye law,

$$C_{ph} = 9NR(T/\theta_D)^3 \int_0^{\theta_D/T} \frac{x^4 e^x}{(e^x - 1)^2} dx \tag{1}$$

(Here, N is the number of atoms per formula unit (n=3, in this case), R is the gas constant, $\theta_D$ is the Debye temperature) at high temperatures. In magnetic materials, one cannot calculate the lattice contribution to total heat capacity easily. One way to measure the lattice heat capacity is to evaluate it at temperatures well above the ordering temperature and to extrapolate it to low temperatures according to the Debye model [9]. Using this method, the Debye temperature has been calculated for this compound and the value is found to be 193 K. This is in the same range as obtained for other members of RTX series [10]. Estimation of electronic heat capacity coefficient in a usual way is not possible in this compound due to the very low transition temperature where the electronic and magnetic contributions dominate.

The estimation of magnetocaloric effect has been done from magnetization (M-H-T) data. Isothermal magnetic entropy change has been calculated utilizing Maxwell's relation

$$\Delta S_M = \int_0^H \left[\frac{\partial M}{\partial T}\right]_H dH \tag{2}$$

In practice, magnetization is measured at discrete fields and temperature interval, for which the change in magnetic entropy can be calculated by the relation [11]

$$\Delta S_M = \sum_i \frac{M_{i+1} - M_i}{T_{i+1} - T_i} \Delta H_i \qquad (3)$$

Here $M_{i+1}$ and $M_i$ are the magnetization values at temperatures $T_{i+1}$ and $T_i$ in field $H_i$, respectively. Fig. 5 depicts the temperature dependence of $-\Delta S_M$ for various field changes. The peak in $-\Delta S_M$ is around the magnetic transition temperature, which increase with increase in field and attains a maximum value of 11.1 J/kg K for a field change of 50 kOe. It is well known that ferromagnetic compounds show positive MCE while antiferromagnetic materials show inverse or negative MCE. From Fig. 5, one can see that the MCE is positive in the whole temperature regime and even for field changes as low as 10 kOe, which confirms that antiferromagnetic interaction is weak in this compound. The value of $-\Delta S_M$ for field change of 50 kOe is comparable to that of other members of RTX family as well as some well known rare earth compounds [11-15].

Recently, it was found that magnetic materials with second order phase transition follow some universal behaviour as far as the MCE variation is concerned. The field dependence of $\Delta S_M^{max}$ has been studied in detail and found to follow $h^{2/3}$ law [16-22]. The field dependence of $\Delta S_M^{max}$ within the framework of molecular field approximation [16] can be expressed in the form of equation given as [17]

$$\Delta S_M^{max} \approx -1.07 qR \left( \frac{g\mu_B JH}{k_B T_C} \right)^{2/3} \qquad (4)$$

Where q is the no. of magnetic ions per formula unit, R is the gas constant and g is Lande-g factor. According to eq. (4), the $\Delta S_M^{max}$ vs. $H^{2/3}$ plot should pass through the origin with $\Delta S_M^{max} = 0$ for H=0, but it has been observed that this plot does not pass through origin and has a intercept [17]. Based on the renormalization group approach to scaling [23], Dong et al. [17] deduced an equation relating $\Delta S_M^{max}$ with H, which is as follows

$$\Delta S_M^{max} = h^{2/3} S(0,1) - S(0,0) \qquad (5)$$

Here $S(0,0)$ is a reference parameter, which must not be zero and h is the reduced field given by $\mu_B H/k_B T_C$. It has been reported that $S(0,1)$ is connected with the spontaneous magnetization at 0

K, i.e., $M_s(0)$. Thus, based on experimental results, the equation (5) has been modified to give the following equation

$$\Delta S_M^{max} = kM_s(0)h^{2/3} - S(0,0) \quad (6)$$

Where, k is a constant. To see the field dependence of $\Delta S_M^{max}$, we have fitted equation (6) by taking $T_N$ in place of $T_C$ and found that the equation fits well with the experimental data. The linear dependence of $\Delta S_M^{max}$ on $h^{2/3}$ implies the strong localization of moments [24]. The order of magnetic phase transition can be distinguished by the parameter $S(0,0)$. A negative sign of $S(0,0)$ indicates second order phase transition, while a positive sign indicates first order phase transition. The $S(0,0)$ calculated from the fit was found to be negative, which confirms the second order magnetic transition in NdCuSi.

To know the origin of MCE, Amaral et al. [25,26] have suggested a theoretical model based on the magnetoelastic coupling and the electron interaction energy using Landau theory of phase transitions. With the help of this model, the temperature dependence of magnetic entropy change has been estimated for NdCuSi. The Gibbs free energy, G(T,M) in terms of magnetization can be written as [27]

$$G(T,M) = G_0 + \frac{1}{2}AM^2 + \frac{1}{4}BM^4 + \frac{1}{6}CM^6 + .... - HM, \quad (7)$$

Where, A, B and C are Landau coefficients, which are temperature dependent. These coefficients are calculated from H/M vs. $M^2$ plots for different temperatures. For minimum energy, the magnetic equation of state can be derived as

$$\frac{H}{M} = A(T) + B(T)M^2 + C(T)M^4. \quad (8)$$

The corresponding magnetic entropy can be obtained by differentiating the magnetic part of G(T,M) with respect to temperature.

$$S_M(T,H) = -\frac{1}{2}A'(T)M^2 - \frac{1}{4}B'(T)M^4 - \frac{1}{6}C'(T)M^6. \quad (9)$$

Where, A', B' and C' are temperature derivative of Landau coefficients. Using the values of A', B' and C', the change in the magnetic entropy can be estimated from the equation given as

$$\Delta S_M(T,H) = S_M(T,H) - S_M(T,0)$$
$$= -\frac{1}{2}A'(T)(M^2 - M_0^2) - \frac{1}{4}B'(T)(M^4 - M_0^4) - \frac{1}{6}C'(T)(M^6 - M_0^6). \quad (10)$$

Fig. 6 depicts the temperature dependence of $-\Delta S_M$ for 50 kOe. The data obtained from the experiment is shown by circles, whereas the entropy change calculated using the Landau model is shown by the solid line. It is clear from Fig. 6 that both these are in good agreement with each other.

The resistivity study of a magnetic material also plays an important role in understanding the magnetic properties because the resistivity is very sensitive to the change of magnetic state. In the case of NdCuSi, the resistivity measurement was done in zero field and in the temperature range of 2-250 K. Fig. 7 shows temperature dependence of resistivity data. It is clear from this figure that this compound shows linear resistivity in the paramagnetic regime.

The inset in Fig. 7 shows an expanded plot at low temperatures. One can see that the slope of the plot changes at about 3.2 K, which arises due the the onset of antiferromagnetic ordering. One point to be noted from the inset is that the resistivity increases near the ordering temperature and then decreases with decrease in temperature. Generally, there is a decrease in the resistivity at the magnetic transition regime due to loss of randomness of magnetic spins in ordered regime. The increase in the resistivity around transition temperature may arise due to two factors (i) carrier scattering by critical spin fluctuations (ii) formation of an antiferromagnetic super zone gaps, which arises due to new Brillouin zone boundaries as a consequence of additional magnetic periodicity [28]. It has been observed that generally superzone gap effect occurs at or below $T_N$ but the present compound shows an increase in resistivity before the Neel ordering, which rules out the presence of magnetic superzone gap effects in the compound [28-30]. Thus, the increase in the zero field resistivity above Neel temperature can be attributed to the carrier scattering by critical spin fluctuations.

**Conclusions**

The NdCuSi crystallizes in the hexagonal crystal structure and shows weak antiferromagnetic behavior, which has been confirmed by magnetization, heat capacity, and magnetocaloric properties. The compound shows considerable MCE at 50 kOe. The maximum entropy change shows $h^{2/3}$ (h is reduced field) dependence with negative S(0,0) value, which indicates the second order of magnetic transition. Moreover, the theoretically calculated MCE using Landau theory is found to compare well with the MCE estimated directly from the M-H-T data. Temperature dependence of electrical resistivity shows a pronounced anomaly at the ordering temperature, which is attributed to the spin fluctuations.

**Acknowledgment**

Sachin Gupta acknowledges CSIR, New Delhi for granting the fellowship. The authors thank D. Buddhikot for his help in resistivity measurement.

org/10.1016/j.mmm.2013.04.004.

**Figure Captions**

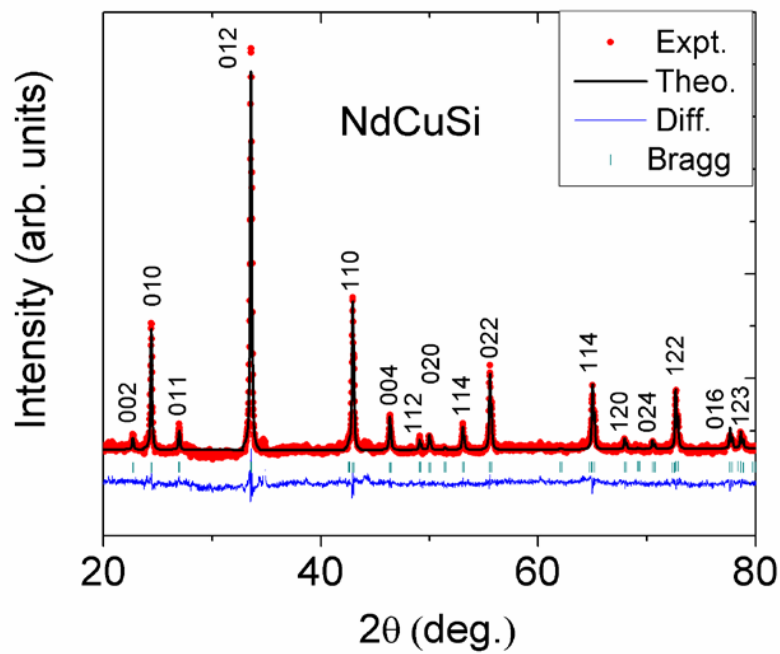

FIG. 1. Room temperature powder XRD pattern for NdCuSi. All the peaks are indexed with their (h k l) values. The bottom plot shows the difference between experimentally observed and theoretical calculated patterns.

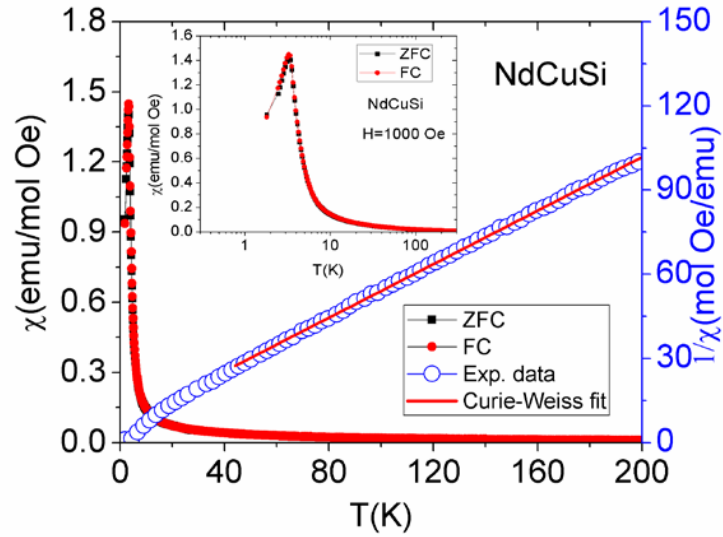

FIG. 2. Temperature dependence of dc magnetic susceptibility (left-hand scale) and the inverse magnetic susceptibility (right-hand scale). Solid line in the inverse susceptibility plot shows the Curie-Weiss fit. The inset shows the semi-log plot of the magnetic susceptibility.

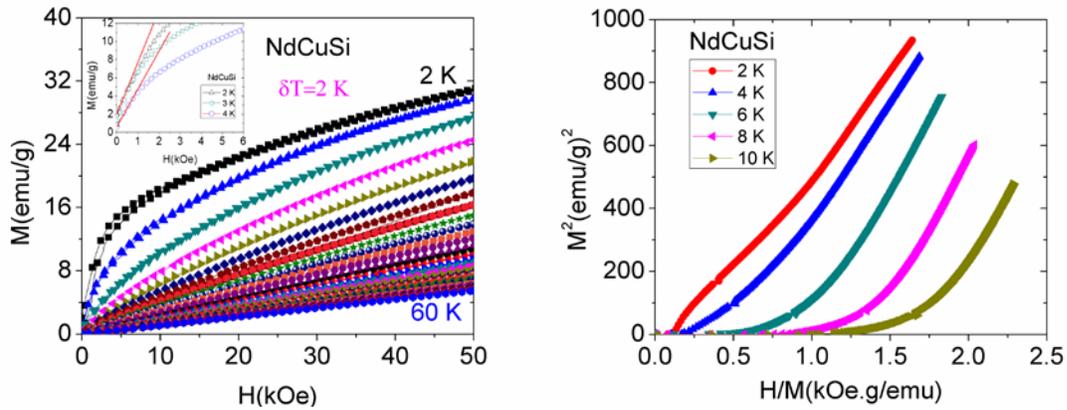

FIG.3. (a) M vs. H plots at different temperatures in NdCuSi (b) The Arrott plots at some selected temperatures. The inset in (a) shows the magnetization at low temperatures and linear fit to it.

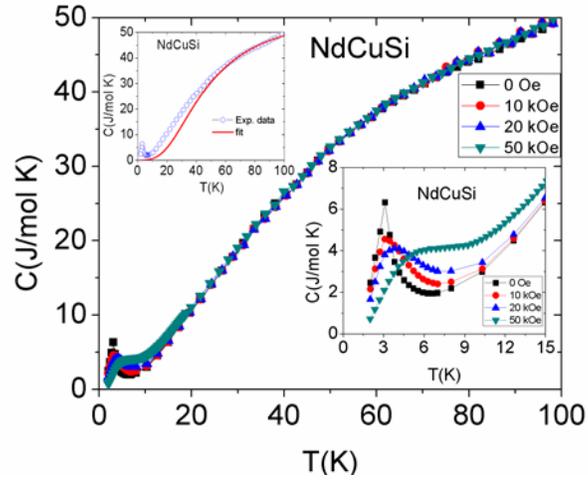

FIG. 4. Temperature dependence of heat capacity at 0, 10, 20, 50 kOe. The upper inset shows Debye fit to zero field heat capacity data and the lower inset shows an expanded plot of heat capacity at low temperatures.

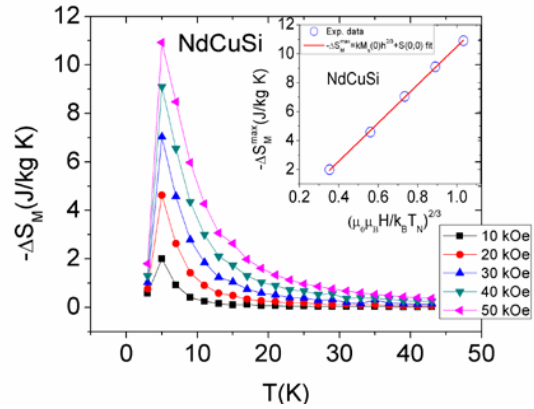

FIG. 5. Temperature dependence of $-\Delta S_M$ at various fields for NdCuSi. The inset shows the field dependence of $\Delta S_M^{max}$.

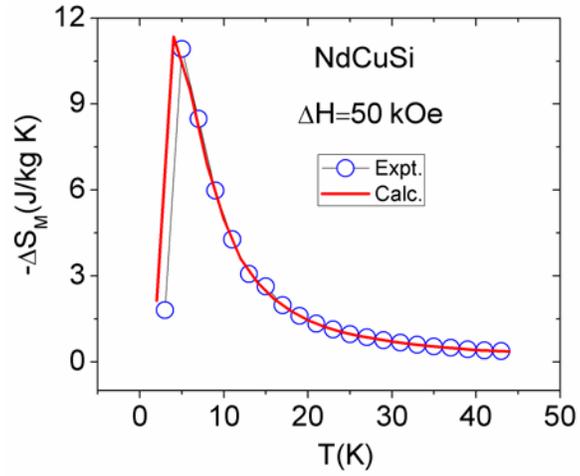

Fig. 6. Temperature dependence of -ΔS$_M$ for 50 kOe in NdCuSi. The solid line shows the calculated -ΔS$_M$ from the Landau theory.

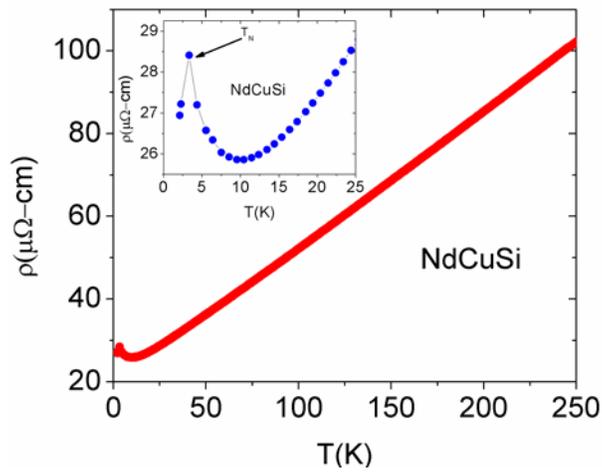

Fig. 7. Temperature dependence of zero field resistivity data for NdCuSi. The inset shows an expanded plot at low temperatures.